\documentclass{ws-p9-75x6-50}

\begin{document}

\title{Hydrodynamical Study of Advective Accretion flow around Neutron Stars}

\author{Banibrata Mukhopadhyay}

\address{Theoretical Physics Group, Physical Research Laboratory,
Navrangpura, Ahmedabad-380009, India
}

\maketitle

\abstracts{
Here we study the accretion process around a neutron star, especially
for the cases where shock does form in the accretion disk.
In case of accretion flows around a black hole, close to the
horizon matter is supersonic. On the other hand for the case
of neutron stars and white dwarfs, matter must be subsonic close
to the inner boundary. So the nature of the inflowing matter
around neutron stars and white dwarfs are strictly different
from that around black holes in the inner region of the disk.
Here we discuss a few phenomena and the corresponding solutions of
hydrodynamic equations of matter in an accretion disk around a
slowly rotating neutron star without magnetic field.}

\hskip1cm
\noindent {\bf International Journal of Modern Physics D} (in Press)
\hskip1cm

\section{Introduction}

We study a few aspects of the accretion process
around a neutron star. Chakrabarti \cite{c96a,c96b}
showed steady state solutions of the inflowing matter in 
advective accretion disk and made a
bridge between the accretion and the wind solutions. In 1997, Chakrabarti
and Sahu \cite{cs97} revisited the Bondi type
flows onto a neutron star. In all those studies the central compact
object, namely the neutron star was chosen non-rotating or in the limit
of very slow rotation. Apart from that, it was also assumed to be
weakly magnetised. In this paper, we consider that
the infalling matter possesses some angular momentum and thereby forms
a disk. Hence our accretion process around neutron stars
is more realistic than Bondi type flow.
 
Far away from the central compact object, the flow of the
infalling matter does not
depend on the nature of the central object, that is whether it is a black
hole or a neutron star and therefore the solutions of the accretion disk
are the same. As the matter
comes closer to the central object, the inner boundary condition becomes
important. In case of a neutron star, matter has to stop at the stellar
surface, i.e., in the very inner region, speed
of the flow must be subsonic. On the other hand if it is a
black hole, due to the absence of a hard surface, in the inner region of
the accretion disk the matter falls attaining
a supersonic speed. This is the basic difference
between the accretion flows around a black hole and a neutron star.
 
When the matter falls towards the compact object, the shock wave may form
\cite{c96a,cs97,c89}. There are several kinds
of shock formation, namely, Rankine-Hugoniot, isentropic and isothermal \cite{landau,c90}.
To form the shock wave in a flow, certain
conditions have to be satisfied at the shock location according to the
nature of the shock. In all kinds of
shock formation, mass flux and total momentum of the matter are always
conserved at the shock location. But there must be a jump in
velocity ($\vartheta$) of the matter. For the Rankine-Hugoniot
shock, at the shock location, matter jumps discontinuously from the
supersonic branch to the subsonic branch generating entropy, keeping
constant energy $E$ of the flow. In the case of isentropic shock,
the discontinuous parameter is energy at the shock location.
In both the cases, temperature jumps up at the shock location.
For the isothermal shock, because of strong radiative cooling, both energy
and entropy jump discontinuously at the shock location.
In this paper, with other discussions we will discuss about the formation
of a shock in the accretion disk mainly around a neutron star. We will also
study the general behaviour of the matter flow around a neutron star with or without
shock and how it differs from that around a black hole.

Around a black hole only one stable shock formation is possible. There, after forming
the shock, matter comes down from the supersonic branch to the subsonic branch and passing
through the inner sonic point falls into the black hole. In the
case of a neutron star, the formation of the shock in a particular matter
flow (upto stellar surface) may happen twice in certain situations.
For a neutron star, after the formation of
a shock (exactly in the same way as for the black hole accretion), matter becomes subsonic.
Then passing through the inner sonic point it attains supersonic speed.
As close to the stellar surface the matter speed must be subsonic, another
shock formation is necessary if the matter has to fall
on the surface of the compact star. Thus, around a compact star with hard
surface , if one shock forms, another shock has to form in the flow.
 
To study all the above mentioned properties of the infalling matter
towards the compact object, we consider viscous flows with angular
momentum. In the next section, we briefly discuss about the basic equations
of the problem. In {\S 3} we indicate the solution procedure and
discuss about the possible solutions. In {\S 4} we show solutions
and finally in {\S 5} draw our conclusions.

\section{Basic Equations of the Flow}

We use a well understood model of the advective
accretion flow close to the black hole and the neutron star in the sub-Keplerian region
of the disk \cite{c96a}.
According to our assumption, viscosity alone is responsible for the energy
dissipation in the disk. We solve the
equations given below to obtain the thermodynamic quantities.
 
\noindent (a) The continuity equation:
 
$$
\frac{d}{dx} (\Sigma x \vartheta) =0 ,
\eqno{(1a)}
$$
where, $\Sigma=h(x)\rho$ is the vertically integrated density.
 
\noindent (b) The radial momentum equation:
 
$$
\vartheta \frac{d\vartheta}{dx} +\frac{1}{\rho}\frac{dP}{dx}
+\frac {\lambda_{Kep}^2-\lambda^2}{x^3}=0.
\eqno{(1b)}
$$
Here, the first term represents advection
which actually gives the information of kinetic energy of the infalling
matter. The second term is due to the matter pressure
and then last one is the combination of gravitational
and centrifugal force terms.
 
\noindent (c) The azimuthal momentum equation:
$$
\vartheta\frac{d \lambda(x)}{dx} -\frac {1}{\Sigma x}\frac{d}{dx}
(x^2 W_{x\phi}) =0 ,
\eqno{(1c)}
$$
with $W_{x\phi}$ is the vertically integrated viscous stress giving rise
to the azimuthal pressure.
 
\noindent (d) The entropy equation:
 
$$
\Sigma v T \frac{ds}{dx} = \frac{h(x) \vartheta}{\Gamma_3 - 1}
\left(\frac{dp}{dx} - \Gamma_1 \frac{p}{\rho}\right)=Q^+-Q^-
$$
$$
\hskip-0.95cm = f(\alpha, x, {\dot m}) Q^+ .
\eqno{(1d)}
$$
Here, $Q^+$ and $Q^-$ are the viscous heat gained
and lost by the flow respectively where for simplicity $Q^-$ is chosen proportional
to $Q^+$ with proportionality constant $g=
1-f(\alpha, x, \dot{m})$, where $f(\alpha, x, \dot{m})$
is the cooling factor and ${\dot m}$
is the mass accretion rate in unit of the Eddington rate.
The cooling is considered to be provided by bremsstrahlung and
Comptonization effect. We use the standard definitions
of $\Gamma$ \cite{cox},
$$
\Gamma_3=1+\frac{\Gamma_1-\beta}{4-3\beta},
\eqno{(2a)}
$$
$$
\Gamma_1=\beta + \frac{(4-3\beta)^2 (\gamma -1 )}{\beta + 12 (\gamma -1)(1-\beta)}
\eqno{(2b)}
$$
and $\beta (x)$ is the ratio of gas pressure to total pressure as,
$$
\beta(x) = \frac {\rho k T/\mu m_p}{\rho k T/\mu m_p + {\bar a} T^4/3 }.
\eqno{(3)}
$$
Here, ${\bar a}$ is the Stefan's constant, $k$ is the Boltzmann constant, $m_p$
is the mass of the proton, $\mu$ is the mean molecular weight.
Using the above definitions, eqn. (1d) becomes,
$$
\frac{4-3\beta}{\Gamma_1-\beta} \left[\frac{1}{T}\frac{dT}{dx}
-\frac{1}{\beta}\frac{ d \beta}{dx} -
\frac{\Gamma_3 - 1}{\rho}\frac{d\rho}{dx} \right]
$$
$$
 = f(\alpha, x, {\dot m}) \frac{Q^+}{{\vartheta}Ph(x)}=
\frac{f{\alpha}x}{v}\frac{d{\Omega}}{dx}.
\eqno{(1e)}
$$
Here, we concentrate on the solutions with constant $\beta$.
Actually, we study the relativistic flows where
$\beta\sim 0$ and consequently $\Gamma_1=\Gamma_3=\frac{4}{3}$.
Although very far away from the compact object, flows need not be
relativistic (hence $\beta$ need not be $\frac{4}{3}$), here for simplicity,
we consider $\beta$ as a constant throughout the particular cases. Similarly,
we consider the cases for $f(\alpha, x, {\dot m})$ = constant, though
it is easy to understand that $f\sim 0$ in the Keplerian disk region and is
greater than $0$ near the compact object depending on the
efficiency of cooling (governed by ${\dot m}$, for instance).
We use the Paczy\'nski-Wiita pseudo-potential \cite{pw}
to describe the geometry of space using non-relativistic equations
(as $\frac{\lambda^2_{Kep}}{x^3}=\frac{1}{2(x-1)^2}$). The half-thickness of the disk
$h(x)\sim ax^{1/2}(x-1)$ at a radial
distance $x$ is obtained from vertical equilibrium assumption,
where $a$ is the adiabatic sound speed ($a^2={\gamma}P/\rho$),
$\lambda(x)$ is the specific angular momentum,
$\vartheta$ is the radial velocity, $s$ is the entropy density
of the flow. The constant $\alpha$ above is the Shakura-Sunyaev 
\cite{ss} viscosity parameter used to express stress tensor in terms of the total
pressure $\Pi$ due to the radial motion. We follow Chakrabarti \cite{c96a,c96b}
to do the calculations for total pressure,
stress tensor $W_{x\phi}$, viscous heat generation which are
not repeated here.
From the continuity equation (Eqn. (1a)), we find the mass accretion rate to
be given by
$$
\dot{m}=2\pi\rho{h(x)}{\vartheta}x.
\eqno{(4)}
$$
Here, $2\pi$ is the geometric factor. The radial co-ordinate $x$ is
expressed in unit of Schwarzschild radius and the velocity is in unit
of speed of light.
 
From the azimuthal momentum Eqn. (1c) we get,
$$
\lambda-\lambda_{in}=\alpha\frac{x}{\vartheta}a^2\left(\frac{2}{3\gamma-1}+
M^2\right),
\eqno{(5)}
$$
where, $M=\vartheta/a$ is the Mach number of the flow and $\lambda_{in}$ is
the specific angular momentum at the inner edge of the flow.
 
\section{Solution procedure and Discussion}
 
From the given set of Eqns. 1(a-c,e) one can eliminate $\frac{da}{dx}$ and
$\frac{d\rho}{dx}$ terms and finally obtain the equation of the form
$$
\frac{d\vartheta}{dx}=\frac{F_1(x,a,\vartheta)}{F_2(\vartheta,a)},
\eqno{(6)}
$$
where,
$$
F_1=\left(\frac{\gamma+1}{a(\gamma-1)}-\frac{4a\alpha^2 f}{\vartheta^2
(3\gamma-1)}\left(\frac{2}{3\gamma-1}+\frac{\vartheta^2}{a^2}\right)\right)
\left(a\frac{5x-3}{2x(x-1)}+\frac{\gamma}{a}\left(\frac{l^2}{x^3}-
\frac{1}{2(x-1)^2}\right)\right)
$$
$$
+\frac{5x-3}{2x(x-1)}-\frac{\alpha f}{\vartheta}
\left(\frac{2}{3\gamma-1}+\frac{\vartheta^2}{a^2}\right)\left(\alpha \vartheta
\left(\frac{2}{3\gamma-1}\frac{a^2}{\vartheta^2}+1\right)-\frac{2\lambda^2}{x^2}\right),
$$
$$
F_2=\left(\frac{\gamma+1}{a(\gamma-1)}-\frac{4a\alpha^2 f}{\vartheta^2
(3\gamma-1)}\left(\frac{2}{3\gamma-1}+\frac{\vartheta^2}{a^2}\right)\right)
\left(-\frac{\gamma}{a}\left(\vartheta-\frac{a^2}{\gamma \vartheta}\right)
\right)+\frac{1}{\vartheta}
$$
$$
-\frac{\alpha^2 f}{\vartheta}\left(\frac{2}{3\gamma-1}+
\frac{\vartheta^2}{a^2}\right)\left(1-\frac{2a^2}{(3\gamma-1)\vartheta^2}\right).
$$
We shall concentrate upon the sub-Keplerian flow starting from the region where
the flow deviates from Keplerian to sub-Keplerian. Here the flow
could be of two kinds.  In one case matter does not
attain supersonic speed at all, thus there is no sonic point in this
flow. In another case, matter passes through the sonic point and
attains supersonic speed, then it takes again the subsonic
branch (that is explained in later section) and falls onto the stellar surface.
 
For a complete run, we supply the basic parameters, namely, the location of
the sonic point $x_{cr}$, the specific angular momentum at the inner edge
of the flow $\lambda_{in}$, the polytropic index $\gamma$, the ratio $f$ of advected
viscous heat flux $Q^+-Q^-$ to heat generation rate $Q^+$, the viscosity parameter
$\alpha$, the accretion rate ${\dot m}$ and the mass of the compact object
in unit of solar mass. The derived quantities
are: $x_K$ where the Keplerian flow becomes sub-Keplerian,
the ion temperature $T$, the flow density $\rho$, the radial velocity
$\vartheta$ and the azimuthal momentum $\lambda$ of the entire flow from $x_K$ to the
stellar surface. Far away from a compact object, the matter speed is very low
(much lower than the sound speed), on the other hand, close to the compact
object the matter speed may overcome the speed of sound. Thus, in that case,
there is a location, namely {\it sonic point}, where this transition
of the matter speed from the subsonic to
the supersonic occurs. From the expression of $F_2$, it is very clear that at that
location it is zero. Therefore, to have a continuous velocity gradient,
$F_1$ must be zero at that radius.  Thus, in the flow, if sonic points exist,
at the sonic location both the numerator
and denominator of $\frac{d\vartheta}{dx}$ vanish and giving two equations,\\
$F_1(x,a,\vartheta)=0$ and $F_2(\vartheta,a)=0$.\\
The equation $F_2(\vartheta,a)=0$ simply gives the expression for
Mach number (M) at the sonic point.
As $F_1(x,a,\vartheta)=0$ is quartic equation, in
principle there will be four roots of $x$ of which either all are real or
two real and two complex or all are complex. In a particular flow,
if real roots at all exist, first it should be checked whether its location
is inside the `$x=1$' surface (according to the Paczy\'nski-Wiita potential)
or not. If the root $x>1$, sonic transition is possible for that
flow. If there are two real physical roots, two possible
sonic transitions exist. In earlier studies of the black hole
accretion disk, it is seen that for existence of a single sonic point
outside the horizon there is no shock formation \cite{c96a}. After passing
through that (inner) sonic point matter falls into the black hole. If
there are two physical sonic points, first, passing through the outer sonic
point matter will attain supersonic speed. Then if the shock condition
is satisfied matter will jump from supersonic to subsonic branch
and after passing through the inner sonic point it will fall into
the black hole. On the other hand if the roots are complex then there is no solution of
accretion flow for that particular parameter region having
sonic point. But, there may be a solution of the
accretion disk around neutron star without having any sonic point,
as we have mentioned above that this is one kind of solution around the neutron star.
There may be an another case, where after passing through the inner
sonic point at the closer region to the compact star matter attains
supersonic speed. If shock forms
in the flow, at the shock location matter jumps from supersonic branch to
subsonic branch and becoming subsonic falls onto the surface of
neutron star. From the
early works of Chakrabarti \cite{c96a,c96b,c89,c90} and numerical
simulations done by Molteni et al. \cite{mlc,mrc,msc} and Sponholz \& Molteni 
\cite{sm} it was concluded that in the accretion disk around a black hole
only one shock is possible. The formation of such a shock in accretion disk
was also verified and studied by other independent groups 
\cite{nabuta,yang,lu97a,lu97b}.
Recently Das et al. \cite{dcc01} have analysed this shock in accretion disk
analytically. The formation of shock strictly depends
on whether the flow satisfies the shock conditions or not.
If the Rankine-Hugoniot conditions are satisfied by the flow, matter
always likes to jump from supersonic to subsonic branch which is of
higher entropy and hence the more stable branch at the same energy. The shock conditions are
given below as \cite{c89}:\\
\noindent (a) conservation of energy at the shock location
$$
E_+=E_-,
\eqno{(7a)}
$$
\noindent (b) conservation of the mass flux at the shock
$$
{\dot m}_+={\dot m}_-,
\eqno{(7b)}
$$
\noindent (c) momentum balance condition
$$
W_++\Sigma_+ \vartheta^2_+=W_-+\Sigma_- \vartheta^2_-
\eqno{(7c)}
$$
and\\
\noindent (d) generation of entropy
$$
s_+>s_-, \hskip0.5cm T_+>T_-.
\eqno{(7d)}
$$
Here, subscript `$+$' and `$-$' refer for quantity just after and before
the shock respectively and
$W$ denotes the vertically integrated pressure as $W=\int_{-h}^{+h}Pdx=
2P I_{n+1}h$, where $I_n=\frac{(2^n n!)^2}{(2n+1)!}$, $n$ being the polytropic
index and $P$ is the pressure at the equatorial plane.
Whenever the Rankine-Hugoniot shock forms and
matter jumps from supersonic to subsonic branch, these four conditions strictly
satisfy. Alternatively, we can say whenever the inflowing matter satisfies
these four conditions, shock forms in the flow.
In this present paper we will show
that in the disk around a neutron star, formation of two shocks is
very natural. Thus in a certain parameter region matter may change
its branch (i.e., it will appear as transonic) twice in a
particular flow. Sometimes it may happen that shock does not
form in a flow when matter attains supersonic
speed at far away ($\sim 50 - 90$ Schwarzschild radius)
and continues to fall smoothly towards
a black hole. This case is unstable, any disturbance created
into the flow may cause a change of the matter from the supersonic
to the subsonic branch via shock. For these unstable cases, at the very inner
region of the disk (much inside than the radii $\sim 10-15$ where
usually shock forms around black hole \cite{c96a,mc00})
still a shock may form and the matter may jump
from the supersonic branch to the subsonic branch and fall onto stellar surface.
Here, it has to be noted that the location, where the shock conditions
(here, according to Rankine-Hugoniot shock) satisfy for a
particular flow, must be greater than the outer radius of the neutron
star, otherwise there is no scope to form the shock.
There is an another category of matter flow with shock. There,
at around $10-15$ Schwarzschild radius one shock forms in a
similar way as the shock formation in the black hole
accretion disk. Then after the matter passes through the inner sonic point
close to the neutron star's surface a second shock forms and becoming
subsonic the matter falls onto the stellar surface (again the
necessary condition is the outer radius of the star must be shorter
than the shock location). When four real roots exist, it is seen that
for any physical parameter set one of the root is always inside the
horizon of the black hole or the outer surface of the star ($x=1$) \cite{c89,c90}.
Out of the other three roots, two are of the `X' type and
other is of the `O' type sonic point. We know that the `O' type sonic point
is not physical. In the case of a shock, sonic transitions occur through
the `X' type
sonic points which are either side of the `O' type. So altogether
three different cases might be occurred around a neutron star, they are:
(1) no shock formation, (2) unstable one shock formation,
(3) two shocks formation.
As because, at most two physical sonic points may exist, maximum two shocks
may occur in accretion flow around a neutron star.
\begin{figure}
\vbox{
\vskip -5.8cm
\centerline{
\psfig{file=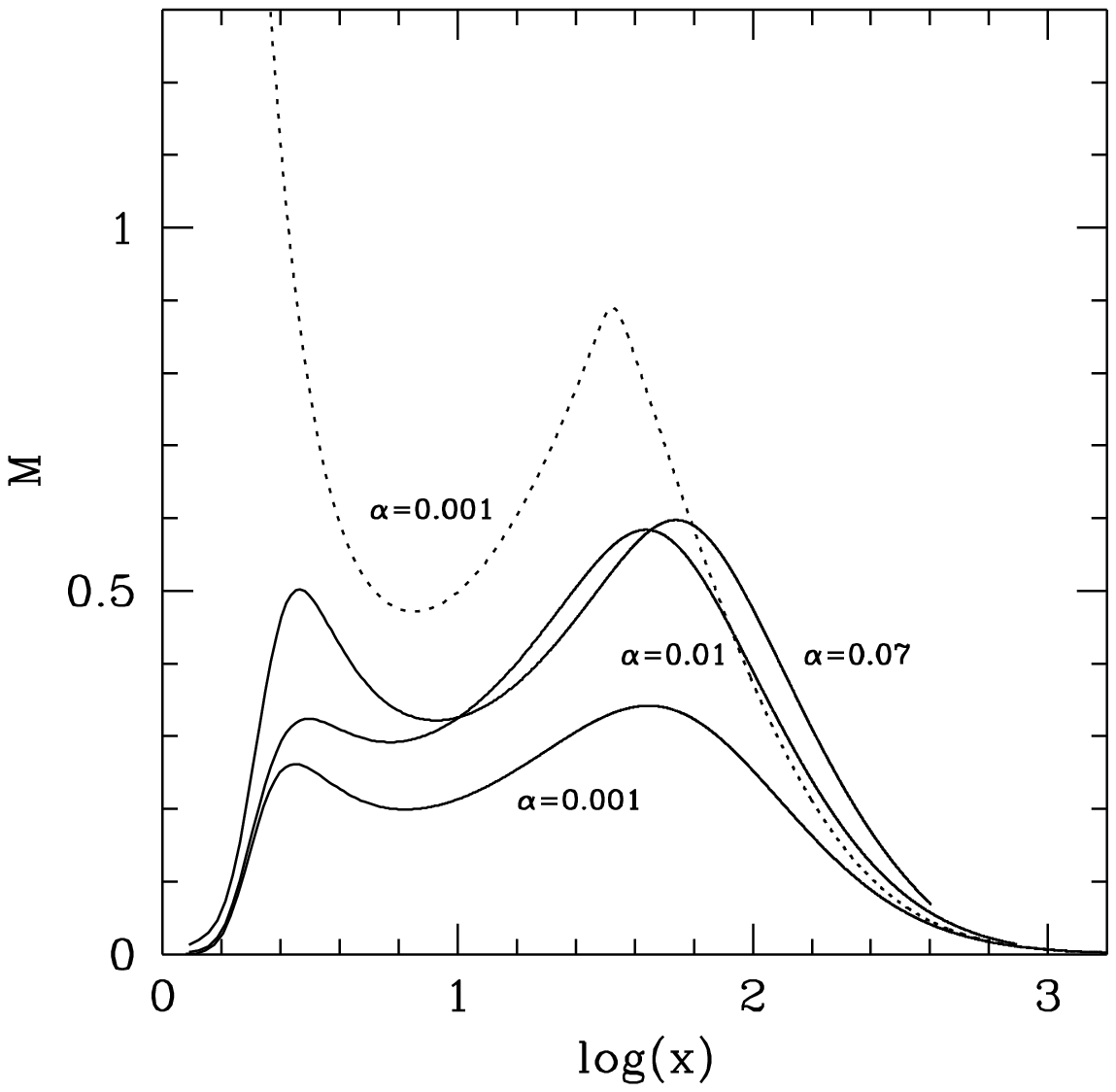,height=15cm,width=10cm}}}
\noindent {{\bf Fig. 1:}
Variation of Mach number $M$ as a function of logarithmic radial distance.
Solid curves are for neutron star and dotted one is drawn for accretion
around black hole. For different values of $\alpha$ (are shown on each
curve) variation changes. For complete set of parameters see TABLE-I.
Here, no shock forms in the flow.
}
\end{figure}
 
\begin{figure}
\vbox{
\vskip -5.8cm
\centerline{
\psfig{file=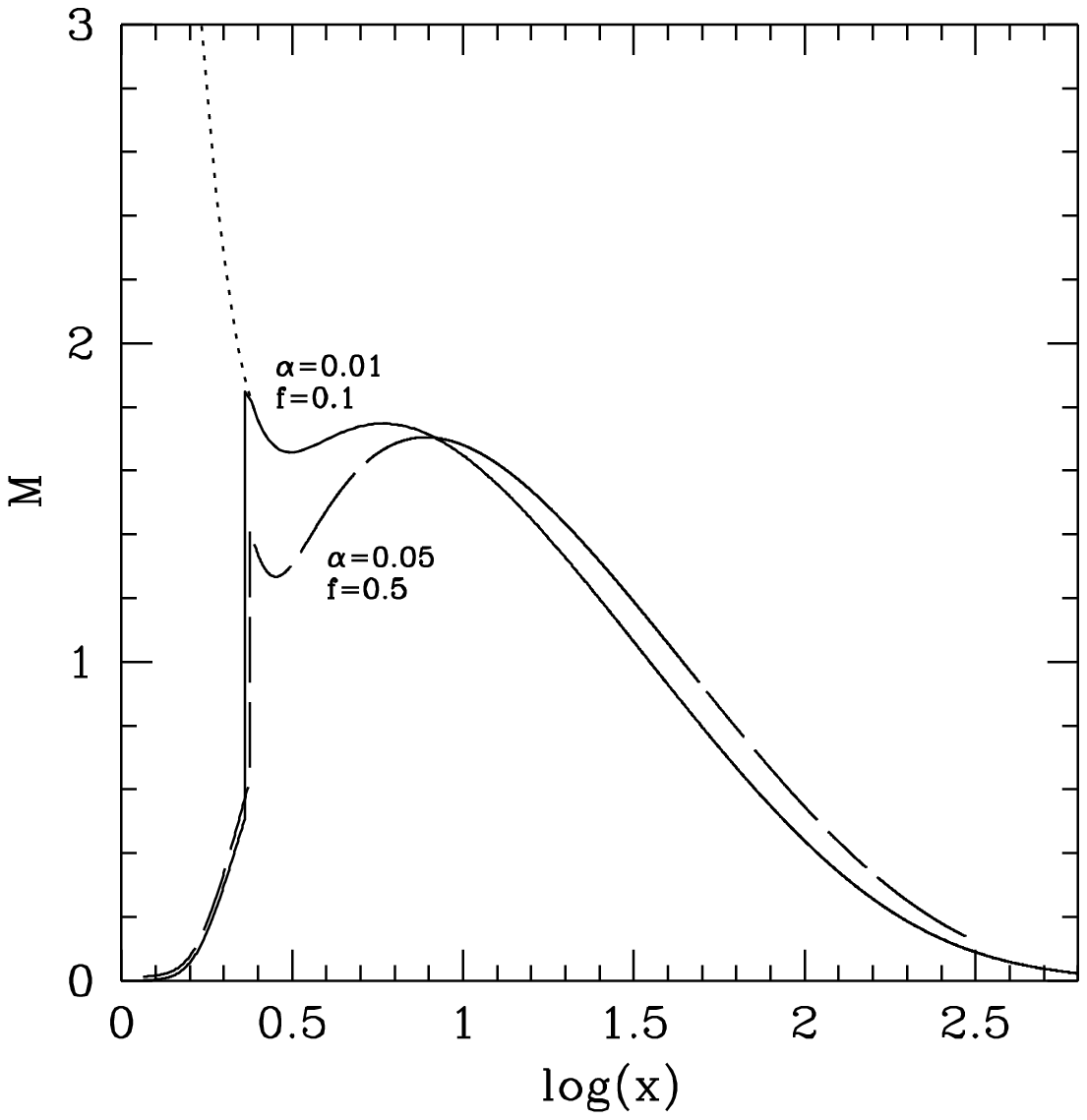,height=15cm,width=10cm}}}
\noindent {{\bf Fig. 2:}
Variation of Mach number $M$ as a function of logarithmic radial distance.
Solid and long dashed curves are for neutron star, dotted curve is drawn
for accretion around black hole. For complete set of parameters see
TABLE-II. Here, one shock is formed in the flow around neutron star.
}
\end{figure}

\section{Results}

We have studied the accretion phenomena around a slowly rotating neutron
star with weak magnetic field for different sets of physical parameters.
Here we choose a few sets of suitable parameters to show the behaviour
of matter flow for some interesting cases. Our main
interest in this paper is to study the dynamical behaviour of the flow in
a steady-state situation. Thus, following the standard practice,
we will concentrate on to study the variation
of the Mach number, from which it is easy to understand the nature of the matter
speed; when it attains the supersonic speed and when it comes down to the
subsonic branch.
 
\subsection{CASE I}
\noindent
Here we choose the parameters in such a manner that the
matter always stays in the subsonic branch. We list the parameters for three
different cases in TABLE-I. We choose same ${\dot m}$, $M_s$ (mass of the
neutron star in the unit of solar mass) and $f$ for all
three cases but different viscosity parameter $\alpha$. We have listed the
values of sonic point $x_{cr}$ for each case. Then we indicate the
number of shock formed by `S' and the corresponding shock location by `$x_s$'.
 
\vskip0.5cm
{\centerline{ TABLE-I}}
\vskip0.5cm
\begin{center}
{\small
\vbox{
\hspace {-1.5cm}
\begin{tabular}{llllllllll}
\hline
\hline
$M_s$ & $\gamma$ & $x_{cr}$ & $\lambda_{in}$ & $\alpha$ & ${\dot m}$ & $f$ & $x_K$ & S & $x_s$ \\
\hline
\hline
$2$ & $4/3$ & NO & $1.6$ & $0.07$ & $1$ & $0.1$ & $401$ & $0$ & NA \\
$2$ & $4/3$ & NO & $1.6$ & $0.01$ & $1$ & $0.1$ & $783.7$ & $0$ & NA \\
$2$ & $4/3$ & NO & $1.65$ & $0.001$ & $1$ & $0.1$ & $1655.7$ & $0$ & NA \\
\hline
\hline
\end{tabular}
}}
\end{center}
\vskip1.0cm
 
It is seen from the TABLE-I that as $\alpha$ decreases $x_K$ increases
as well as residence time of the matter in the disk increases.
In Fig. 1, we show the variation of Mach
number with  different values of $\alpha$. As $\alpha$  decreases, energy momentum
transfer rate decreases, as well as the corresponding Mach number and the
centrifugal force decrease. Thus with the increase of $\alpha$, incoming
matter attains more speed as well as more centrifugal force in the
outward direction (as there is quadratic dependence on velocity). Finally
matter falls on to the surface of neutron star. For the comparison we
also show one solution for the matter (of $\alpha=0.001$) that falls
into a black hole of mass $10M_\odot$. In that case it is clear
from Fig. 1 that passing through the sonic point $x_{cr}=2.7945$,
matter falls supersonically into the black hole.
 
\subsection{CASE II}
 
Here we choose the parameter in such a manner that close to the surface
of the neutron star, shock forms. As we discussed in previous section, this
case is slightly unstable as matter attains a supersonic speed at far
away from the neutron star and continues to fall supersonically from
$x\sim 50$ to just before the shock in the inner region of the disk. Usually in
stable cases shock forms at a much greater radius as we will show in the next
subsection (CASE III). Here, no shock forms at the outer radius and matter
smoothly falls supersonically. But in the very inner region it satisfies the
Rankine-Hugoniot shock conditions and jumps
discontinuously from the supersonic branch to the subsonic branch
and falls on to the star's surface. We list the parameters for two different
cases in TABLE-II. We choose same ${\dot m}$, $M_s$ for both the cases
but different viscosity parameter $\alpha$ and cooling factor $f$. We also
indicate the number of shocks formed (by S) and its location ($x_s$)
as in CASE I. Two sonic points are given as $x_{cr1}$ and
$x_{cr2}$.
 
\vskip0.5cm
{\centerline{ TABLE-II}}
\vskip0.5cm
\begin{center}
{\small
\vbox{
\hspace {-1.5cm}
\begin{tabular}{lllllllllll}
\hline
\hline
$M_s$ & $\gamma$ & $x_{cr1}$ & $x_{cr2}$ & $\lambda_{in}$ & $\alpha$ & ${\dot m}$ & $f$ & $x_K$ & S & $x_s$
\\
\hline
\hline
$2$ & $4/3$ & 50 & 2.869 & $1.6$ & $0.05$ & $1$ & $0.5$ & $481.4$ & $1$ & 2.38 \\
$2$ & $4/3$ & 50 & 3.156 & $1.6$ & $0.01$ & $1$ & $0.1$ & $783.6$ & $1$ & 2.3 \\
\hline
\hline
\end{tabular}
}}
\end{center}
\vskip1.0cm
 
In Fig. 2 we show the variation of Mach number with logarithmic radial
distance. For higher viscosity, matter is slowed down at the centrifugal
pressure dominated region (centrifugal effect is maximum at $x\sim3.2$)
more strongly because of high rate of energy momentum transfer. It can
be noted that if $\alpha$ is very high, say $\ge 0.1$, centrifugal barrier
may not appear because of very small $x_K$.
At that low radius centrifugal force could not be
comparable to the gravitational force, so the centrifugal pressure
is smeared out. In Fig. 2, we see that the shock forms at very close to
the neutron star for both the cases. The shock location is $\sim 2.3-2.4$ (see TABLE-II)
and generally for the neutron star of mass $2M_\odot$, outer radius is smaller
than that shock location \cite{cook,dey}. Thus the inner shock is possible here.
 
\subsection{CASE III}
 
Here we choose the parameter in such a manner that the shock forms twice in
the flow around the neutron star. We have chosen two sets of parameters given in
TABLE-III to describe this case. We choose the same accretion rate
and mass of the star for both the cases. Two shock locations $x_{s1}$ and
$x_{s2}$ are listed
form on either side of the sonic points (given as $x_{cr1}$ and $x_{cr2}$).
 
\vskip0.5cm
{\centerline{ TABLE-III}}
\vskip0.5cm
\begin{center}
{\small
\vbox{
\hspace {1.5cm}
\begin{tabular}{llllllllllll}
\hline
\hline
$M_s$ & $\gamma$ & $x_{cr1}$ & $x_{cr2}$ & $\lambda_{in}$ & $\alpha$ & ${\dot m}$ & $f$ & $x_K$ & S & $x_{s1}$ & $x_{s2}$ \\
\hline
\hline
$2$ & $4/3$ & 50 & 2.911 & $1.6$ & $0.07$ & $1$ & $0.1$ & $401$ & $2$ & 15.012 & 2.8025 \\
$2$ & $4/3$ & 50 & 2.869 & $1.6$ & $0.05$ & $1$ & $0.5$ & $481.4$ & $2$ & 13.9 & 2.737\\
\hline
\hline
\end{tabular}
}}
\end{center}
\vskip1.0cm
 
During its motion, at first the matter satisfies the Rankine-Hugoniot shock conditions
at a outer radius in comparison to the shock location of CASE II.
The formation of the shock at this location is exactly similar to that which could
be formed around a black hole. By this shock formation matter
jumps from its unstable branch to a stable subsonic branch of higher
entropy. This shock formation is independent of the central
compact object, whether it is a black hole or a neutron star. After that,
passing through the inner sonic point matter again attains a
supersonic speed (see Fig. 3). Now another shock may form, depending on the
nature of compact object. If it is a black hole
there is no question of the second shock as the speed of the
matter must be supersonic close to the horizon. On the other hand, if there
is a neutron star, without having another shock matter can not
reach the surface of the star in this present situation.
This second shock occurs exactly in the similar way
as the shock in CASE II, if the corresponding Rankine-Hugoniot shock conditions are
satisfied. In Fig. 3, we show the formation of two shocks in the accretion
flows around a neutron star. We show the Mach number variation of the
accreting matter around neutron star for two different sets of
initial parameter (solid and long-dashed curve). Two shock locations
are denoted by sck1 and sck2. We also show the Mach number variation
for the case if the matter with $\alpha=0.07$ had fallen towards
a black hole of mass $10M_\odot$ by the dotted (stable flow with shock)
and short-dashed (unstable flow without shock) curves. It is clear
from the figure that upto just before the formation of the second shock, the solution
around a black hole and a neutron star are similar; as if the matter does not
know about the nature of central object. Thus, the formation of
two shocks depends upon the nature of the
compact object and the possibility of satisfaction of shock conditions.
For certain choices of the flow parameter values, even for a neutron star
the Rankine-Hugoniot shock conditions
may not satisfy, though the matter attains a supersonic speed. For these
cases, the matter will not reach the neutron star.
Again it should be reminded that the inner shock location must be outside
the radius of the neutron star. Here, as the radius of the shock location
($\sim 2.7-2.8$) is greater than the usual radius of a neutron star of mass
$2M_\odot$ \cite{cook,dey}, inner shock forms safely.
 
Here we also show the temperature variation in an accretion disk around a neutron
star and a black hole in Fig. 4. Solid and long-dashed curves indicate the
temperature profiles around a neutron star (indicated
as NS in the figure) when viscosities of the infalling matter are $0.07$ and
$0.05$ respectively (for other parameters, see TABLE-III). At the shock
locations (as indicated by sck1 and sck2 in figure) temperature rises
discontinuously. By the dotted and short-dashed curves we show the temperature
variation of the inflowing matter of viscosity $0.07$ when it falls
towards a black hole of mass $10M_\odot$ (as indicated BH in the figure).
Dotted curve is drawn for stable shock case and short-dashed curve is
for unstable no-shock case (if the same matter had fallen
without forming a shock). Although the virial temperature of the accretion
disk may be very high as of the order of $10^{11}$K, following Chakrabarti \&
Mukhopadhyay \cite{cm99} and Mukhopadhyay \& Chakrabarti \cite{mc00}
we have taken into account the cooling process so that the temperature
reduces to of the order of $10^9$K. As we consider the entire flow is
relativistic, i.e., $\beta$ is very low (radiation pressure dominated flow),
the soft photon in the disk is very
profuse. Thus, the virial temperature may be reduced by the inverse-Compton
effect. As there is the possibility of the formation of two shocks in the disk
around a neutron star temperature is still high enough. It is very clear from the above discussions
that for the accretion around a neutron star, temperature and density are
higher compared to that around a black hole.
 
\begin{figure}
\vbox{
\vskip -5.8cm
\centerline{
\psfig{figure=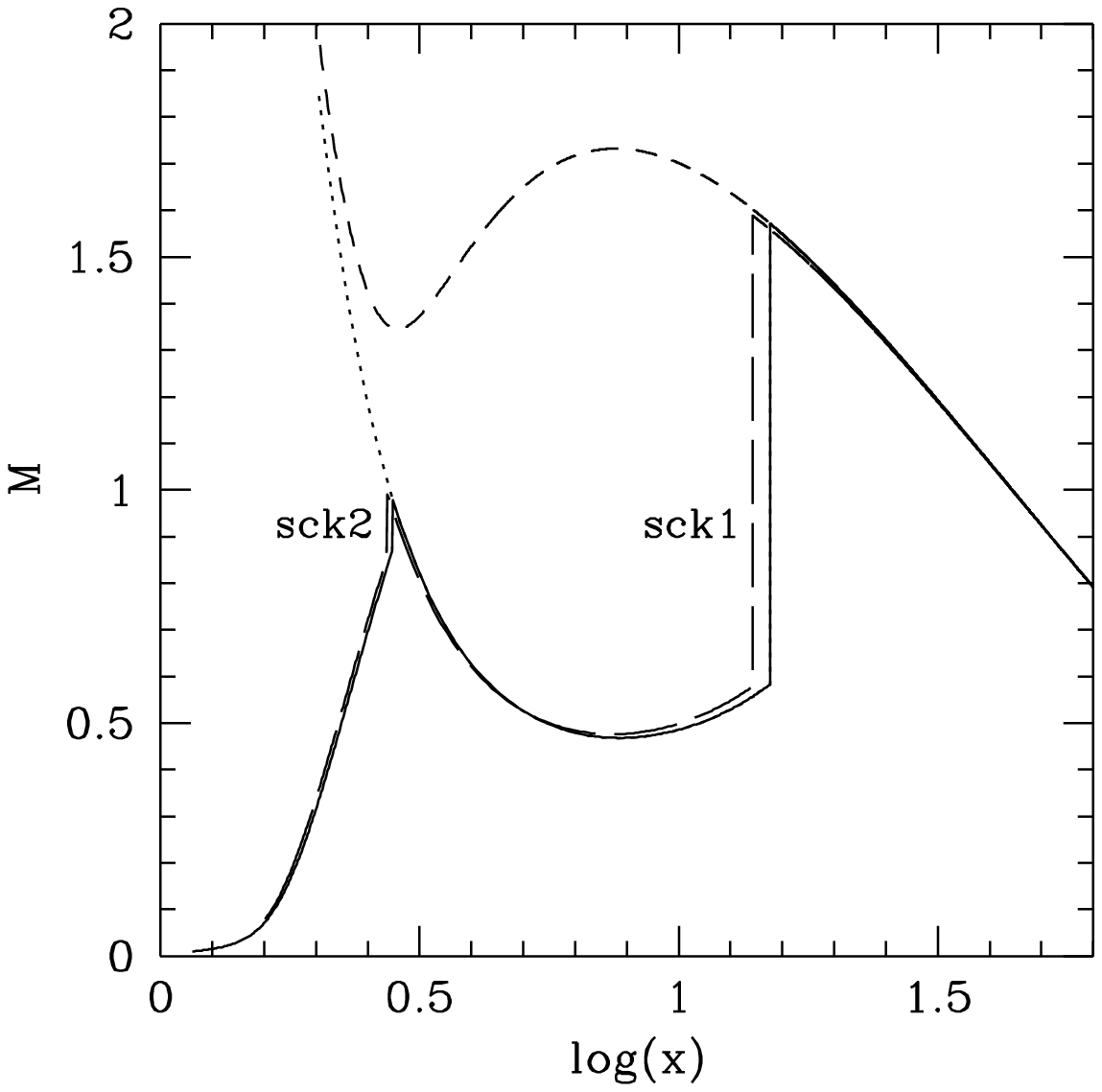,height=15cm,width=10cm}}}
\noindent {{\bf Fig. 3:}
Variation of Mach number $M$ as a function of logarithmic radial distance.
Solid and long-dashed curves are for neutron star and dotted (stable case)
and short-dashed (unstable case) curves are drawn for accretion around
black hole. Clearly two shocks form in the flow around neutron star
which locations are indicated as sck1 and sck2. For complete set of
parameters see TABLE-III.
}
\end{figure}
 
\begin{figure}
\vbox{
\vskip -5.8cm
\hskip 0.0cm
\centerline{
\psfig{figure=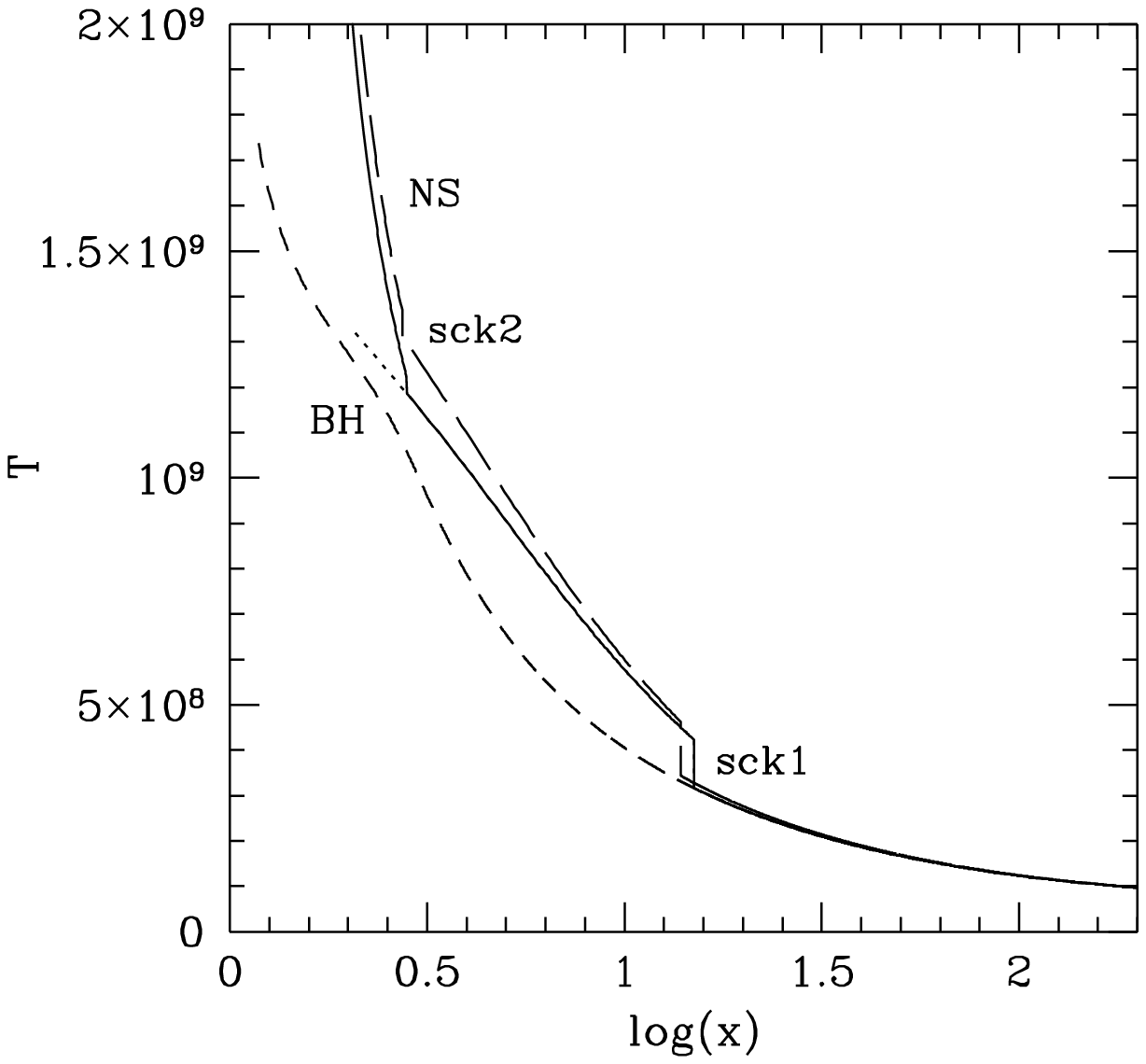,height=15cm,width=10cm}}}
\vskip 0.0cm
\noindent{{\bf Fig. 4:}
Variation of temperature in $K$ as a function of logarithmic radial
distance. By NS and BH we mean the temperature distribution around
neutron star and black hole respectively. We indicate the shock
locations by sck1 and sck2.
}
\end{figure}

\section{Conclusions}

In this paper we study the accretion flows around a neutron star having
very slow rotation and weak magnetic field. We also compare the result
with that around a black hole. We consider the recent concept of the accretion
disk model and study the flow properties. We see that the shock may or may
not form in the disk. We choose three different regions in the parameter space to
show different features of the inflowing matter. We see, though in case
of the accretion disk around a black hole only one shock is possible, around
a neutron star formation of double shock is very natural, if the
outer radius of the neutron star is not very very large (like $\sim 3$
Schwarzschild radius) which is unlikely.
As the incoming matter is slowed
down abruptly at the each shock location, the overall density becomes higher
for a neutron star accretion disk compared to that for a black hole as two shocks
may form around the neutron star. Similarly the temperature in the
disk is comparatively higher, for a neutron star, as the temperature also jumps
at the shock location. The higher density and higher temperature
are not the result only of
shock formation. Even if the shock does not form, the density and the temperature
of the disk are higher close to the neutron star compared to the case
of the black hole, because the matter has to be subsonic there. So, one can
conclude that the accretion disk around a neutron star is very favourable
for nucleosynthesis. Here it can be reminded that all along we have talked
about the sub-Keplerian region of the accretion disk. In this context
we can propose that the study of nucleosynthesis in the sub-Keplerian
accretion disk around a neutron star should be done in future. We know that
the advective accretion disk around a black hole is enough hot for nucleosynthesis
\cite{mc00,cm99,m98,m99,mc01}, which are different from that of the star.
As the density and temperature
might be higher around a neutron star, more efficient nucleosynthesis is expected
close to it. Earlier we saw that the high temperature of accretion disk around
a black hole is very favourable for the photo-dissociations and the proton
capture reactions \cite{mc00}. As the accretion disk around a neutron star is hotter
than that around a black hole, even $^4\!He$ which has a high binding energy
may dissociate into deuterium and then
into proton and neutron. If we consider the accreting matter comes
from the nearby `Sun like' companion star, the initial abundance of
$^4\!He$ in the accreting matter is about $25\%$. Therefore, by the dissociation of
this $^4\!He$, neutron may produce in a large scale which could give rise the
neutron rich elements. Guessoum \& Kazanas \cite{kazan} showed that the profuse
neutron may produce in the accretion disk and through the spallation
reactions $Li$ may be produced in the atmosphere of the star.
When the neutron comes out from the accretion disk by the formation of
an outflow, in that comparatively cold environment
$^7\!Li$ may be produced, which can be detected on the stellar surface.
Earlier it was shown that the metalicity of the galaxy may be influenced
when outflows are formed in the hot accretion disk around black holes \cite{mc00}.
In case of the lighter galaxy, the average abundance of the isotopes of
$Ca$, $Cr$ may significantly change.
Also the abundance of lighter elements, like the isotopes of $C$, $O$,
$Ne$, $Si$ etc. may be increased significantly. As the
temperature of the accretion disk around a neutron star is higher, the expected
change of abundance of these elements and the corresponding influence on
the metalicity of the galaxy is higher.
We would like to pursue all these studies in the next work.
 
Molteni and his coworkers have already shown that the oscillatory nature
of the shock location is related to the cooling and advective time scale
of the matter in the disk \cite{msc,rcm}. Corresponding oscillation frequency is
related to the location of the shock and the observed QPO frequency for the various
black hole candidates can be explained. We know that the QPO
frequencies are observed mainly in two ranges: Hz and KHz \cite{morgan,bulik}.
The frequency depends on the time taken by the
inflowing matter to reach the compact object from the shock location.
From the discussion of Molteni, Sponholtz \& Chakrabarti \cite{msc}, 
it is clear that  as the shock location comes
closer to the compact object QPO frequency increases. According to the
earlier study of the accretion flow around a black hole \cite{c96a}, it was
found that the shock may form at a significantly outer radius. Following
the prescription of Molteni, Sponholtz \& Chakrabarti \cite{msc}, 
we can find the corresponding QPO frequency of the
order of Hz. Those theoretical calculations match with the observed QPO
frequencies, say for the candidates GS 339 4 and 1124 68.
Thus, these lower QPO frequencies can be explained when outer shocks
are formed. Now following the same prescription given by 
Molteni, Sponholtz \& Chakrabarti \cite{msc}, the higher QPO
frequencies can also be explained for the case of a neutron star accretion disk,
because of the inner shock around a neutron star.
As for the presence of an inner shock,  matter takes less time to fall
onto the stellar surface from the shock location, corresponding frequencies will
be higher. Therefore for the explanation of the KHz QPO frequency, the double
shock formation in the accretion disk around a neutron star is very
significant. Our next step of the study will be the theoretical calculation
of the KHz  QPO frequency and to compare it with the observed values.

\section*{Acknowledgments}
The author expresses his gratitude to Professor Sandip K. Chakrabarti for
helpful comments regarding this study.
He is also very grateful to Professor A. R. Prasanna and Dr. Sudip
Bhattacharyya for carefully reading
the manuscript and making their critical comments regarding various
scientific concepts which have been discussed in this paper.

\end{document}